\documentclass[aps,prb,twocolumn]{revtex4}

\usepackage{amsmath}
\usepackage{color}
\usepackage{amssymb}

\usepackage{graphicx}

\begin{document}

\title{ Effect of critical fluctuations on the spin transport in liquid $^3$He }

\author{V.P.Mineev$^{1,2}$}
\affiliation{$^1$Univ. Grenoble Alpes, CEA, IRIG, PHELIQS, GT, F-38000 Grenoble, France\\
$^2$Landau Institute for Theoretical Physics, 142432 Chernogolovka, Russia}

\begin{abstract}
The contribution of pair fluctuations to the spin current in liquid $^3$He in isotropic aerogel 
near the critical temperature of transition to  superfluid state is calculated.
\end{abstract}

\date{\today}
\maketitle

\section{Introduction}
The superfluid state of liquid $^3$He is formed by means the Cooper pairing with spin and orbital angular momentum equal to 1 \cite{Vollhardt2013}. 
Investigation of superfluid phases in high porosity aerogel allows  to study the influence of impurities on superfluidity with 
$p$-pairing \cite{Parpia1995,Halperin1995}. 
In the  normal-state spin diffusion coefficient
of $^3$He in aerogel is determined
both by elastic and inelastic scattering of $^3$He quasiparticles. 
At low temperatures  the collisions between the Fermi liquid quasiparticles induce negligibly small correction to the spin diffusion due to the scattering on aerogel strands \cite{Sauls2005,Dmitriev2015}. 
The field theoretical approach to the calculation of the spin diffusion coefficient in the normal $^3$He in an anisotropic aerogel has been developed in the paper \cite{Mineev2018} in the analogy with  the calculation 
 of electric current in an isotropic metal with randomly distributed impurities performed in \cite{Abrikosov1959}.  Close to the superfluid transition temperature in line with regular spin transport  limited by scattering of quasiparticles on the aerogel there is an additional mechanism determined by the Cooper pairs fluctuations accelerating spin transport  as the critical point is neared. 
 The effects of fluctuations on the thermodynamics and kinetics of a superconductor near the transition point are well known \cite{Varlamov2009}. The theoretical studies of these phenomena have acquired the firm basis since the  publication of the seminal paper by L.G.Aslamazov and A.I.Larkin where the field theoretical approach to the problem has been developed \cite{Aslamazov1968}. The corresponding theory in application to d-wave pairing in layered metals has been developed in the papers
 \cite{Yip1990}. The low temperature (quantum) limit  has been considered in Ref.11.
  In the present paper, I apply 
 this approach  to calculation of the contribution of pair fluctuations to the spin current density above the critical temperature
 in normal $^3$He in isotropic aerogel. 
 
 To make easy comparison with  calculation of paraconductivity in superconductors with $s$-pairing  Ref.9 I begin with definition of the fluctuating propagator for $p$-wave superfluid. An electric current presents a response to the em vector potential. Similarly a spin current is given by  the  response to the nonuniform rotation of the spin space. This allows to perform the derivation of para spin diffusion in liquid $^3$He in  the same spirit as  paraconductivity of a metal near transition to $s$-wave superconducting state.

 \section{Spin diffusion of a fluctuating pair}
 
The order parameter of superfluid phases of $^3$He is given \cite{Vollhardt2013} by the complex $3\times3$ matrix
$ A_{\alpha i} $, where $\alpha$  and $i$ are 
the indices  
numerating the Cooper pair wave function projections  on  spin and orbital axes respectively. 
The second-order term in the Landau free energy density is 
\begin{widetext}
\begin{equation}
F^{(2)}=\left \{\frac{1}{3g}\delta_{rs}- T\sum_{n}\int\frac{d^3p}{(2\pi)^3}G_{{\bf p},\varepsilon_n} G_{-{\bf p},-\varepsilon_n}
\hat p_r\Lambda^s_{{\bf p},\varepsilon_n}
\right \}A_{\alpha r}^\star A_{\alpha s},
\label{fe}
\end{equation}
\end{widetext}
where $g$ is the constant of $p$-wave triplet pairing, $\hat p_r$ is the $r$-component of the momentum unit vector ${\bf p}/|{\bf p}|$.
Here,\begin{equation}
G_{{\bf p},\varepsilon_n}=\frac{1}{i\tilde\varepsilon_n-\xi_{\bf p}}
\label{G}
\end{equation}
 is the normal state quasiparticle Green function and $\Lambda^s({\bf p},\omega)$ is the vertex part. $\xi_{\bf p}=\epsilon_{\bf p}-\mu$ is the quasiparticles energy  counted from the chemical potential, and $\varepsilon_n=\pi T(2n+1)$ are the fermion Matsubara frequences,
$\tilde\varepsilon_n=\varepsilon_n+\frac{1}{2\tau}
{ sign}\varepsilon_n$, $\tau$ is the mean free time scattering of quasiparticles in an isotropic aerogel.  The Planck constant $\hbar$ was everywhere put equal to 1. Correspondingly  the matrix of the fluctuation propagator  is
\begin{widetext}
  \begin{equation}
  L^{rs}({\bf q}, \Omega_k)=\left ( \frac{1}{3g}\delta_{rs}-T\sum_n\int\frac{d^3p}{(2\pi)^3}
  G_{{\bf p},\varepsilon_n}G_{-{\bf p}+{\bf q},\Omega_k-\varepsilon_n}\hat p_r\Lambda^s_{{\bf p},{\bf q}, \varepsilon_{n}}\right )^{-1},
  \label {Fl}
  \end{equation}
  \end{widetext}
where $\Omega_k=2\pi T k$ are the boson Matsubara frequencies.

As it was pointed out in Ref.9 the largest contribution to the conductivity of a fluctuating pair is given by the diagram shown in Fig.1, where wavy lines are the fluctuating propagators, the straight lines are the Green functions and the shaded triangles are the vertex parts.  In contrast to s-pairing, due to  momentum dependent pairing interaction,  all the vertices are not scalar but  vector functions. To find the corresponding analytic expression one must define the spin current.

 The spin current in neutral Fermi liquid can be found  \cite{,Makhlin1992,Volovik1992} as response to  the gradient of angle of rotation of the spin space 
$\mbox{\boldmath$\omega$}_i=\nabla_i\mbox{\boldmath$\theta$}$,
\begin{equation}
{\bf j}_{i}=-\frac{\delta H }{\delta\mbox{\boldmath$\omega$}_i},
\end{equation}
where
\begin{equation}
H=\frac{1}{2m}\int d^3r (D_i^{\alpha\lambda}\psi_\lambda)^\dagger D_i^{\alpha\mu}
\psi_\mu+H_{int},
\end{equation}
\begin{equation}
D_i^{\alpha\beta}=-i\delta_{\alpha\beta}\nabla_i+\frac{1}{2}\mbox{\boldmath$\sigma$}_{\alpha\beta}\mbox{\boldmath$\omega$}_i,
\end{equation}
$\mbox{\boldmath$\sigma$}=(\sigma_x,\sigma_y,\sigma_z)$ are the Pauli matrices,
and $H_{int}$ includes the Fermi liquid interaction and  the interaction with  impurities. 
The response to the gauge field $\mbox{\boldmath$\omega$}_i$ is calculated \cite{Mineev2018} in  analogy with response to the usual vector potential $A_i$ \cite{Abrikosov1959}. 
 The contribution of pair fluctuations to the spin current density above the critical temperature
 corresponding to diagram shown in Fig.1 is
 \begin{widetext}
 \begin{equation}
 {\bf j}_{i}(\omega_\nu)=\int\frac{d^3q}{(2\pi)^3}T\sum_k{\bf B}^{lr}_{i,\alpha\beta}({\bf q}, \Omega_k,\omega_\nu)L^{rs}({\bf q}, \Omega_k)
 ({\bf B}^{st}_{j,\beta\alpha}({\bf q}, \Omega_k,\omega_\nu)\cdot \mbox{\boldmath$\omega$}_j)L^{tl}({\bf q}, \Omega_k+\omega_\nu).
 \label{current}
 \end{equation} 
 Here,  $\omega_\nu=2\pi T \nu$ are the boson Matsubara frequencies,
$L^{rs}({\bf q}, \Omega_k)$ is the fluctuation propagator.
The triangle block 
 \begin{equation}
{\bf B}^{lr}_{i,\alpha\beta}({\bf q}, \Omega_k,\omega_\nu)=T\sum_n\int\frac{d^3p}{(2\pi)^3}v_i\mbox{\boldmath$\sigma$}_{\alpha\beta}\Lambda^l_{{\bf p},{\bf q}, \varepsilon_{n}+\omega_\nu,\Omega_k-\varepsilon_n}
\Lambda^{r\star}_{{\bf p},{\bf q}, \varepsilon_{n},\Omega_k-\varepsilon_n}G_{{\bf p},\varepsilon_{n}+\omega_\nu}
G_{{\bf p}, \varepsilon_{n}}G_{{\bf q}-{\bf p}, \Omega_k-\varepsilon_{n}}
\label {B}
 \end{equation} 
 \end{widetext}
is expressed through  three Green functions (\ref{G})
and the impurity vertex functions $\Lambda^l_{{\bf q}, \varepsilon_{n},\Omega_k-\varepsilon_n}$, $\Lambda^{r\star}_{{\bf q}, \varepsilon_{n},\Omega_k-\varepsilon_n}$. The sign of the complex conjugation in the second  vertex function in Eq.(\ref{B}) corresponds to the opposite direction of the arrows of  the Green function lines in Fig.1 in respect to the first vertex (time inversion).
The impurity vertex functions  are determined by the integral equation
%\begin{widetext}
\begin{eqnarray}
\Lambda^l_{{\bf p},{\bf q}, \varepsilon_{n},\Omega_k-\varepsilon_n}=\hat{ p}_l~~~~~~~~~~~~~~~~~~~~~~\nonumber\\+\frac{1}{2\pi N_o\tau}\int\frac{d^3p}{(2\pi)^3}
G_{{\bf p},\varepsilon_n}G_{-{\bf p}+{\bf q},\Omega_k-\varepsilon_n}\Lambda^l_{{\bf p},{\bf q}, \varepsilon_{n},\Omega_k-\varepsilon_n}.
\label{Lambda}
\end{eqnarray}
%\end{widetext}
Near critical temperature 
the main frequency dependence arises from the fluctuation propagators $L$ having the pole structure. Due to this reason one can neglect by the frequency dependence of the blocks $\bf B$ and the vertices $\Lambda$.  In the integral  Eq.(\ref{current}) are essential only small values of $q$. 
Then, the solution Eq.(\ref{Lambda}) 
is
\begin{equation}
\Lambda^l_{{\bf p},{\bf q}, \varepsilon_{n}}=\hat{p}_l+\frac{iq_lv_Fsign~
\tilde\varepsilon_n}{6|\tilde\varepsilon_n|(1-2|\tilde\varepsilon_n|\tau)}  +\hat q_l{\cal O}(q^3),
\label{Lambda1}
\end{equation}
 here $v_F$ is the Fermi velocity. 
The integral of the product of three Green functions  in linear in respect of wave-vector ${\bf q}$ approximation is
\begin{eqnarray}
N_0\int d\xi G_{{\bf p},\varepsilon_{n}}
G_{{\bf p}, \varepsilon_{n}}G_{{\bf q}-{\bf p},-\varepsilon_{n}}\nonumber\\=-\frac{\pi N_0}{2}\left (i\frac{sign~ \tilde\varepsilon_n}{\tilde\varepsilon_n^2} +\frac{({\bf q}\cdot {\bf v})}{|\tilde\varepsilon_n|^3} \right ),
\label{GGG}
\end{eqnarray}
where $N_0$ is the density of states per one spin projection.
Substituting Eqs.(\ref{Lambda1}), (\ref{GGG}) in the Eq.(\ref {B}) and performing the integration over angles we obtain
\begin{widetext}
\begin{equation}
{\bf B}^{lr}_{i,\alpha\beta}=\frac{1 }{60}\frac{N_0v_F^2}{(2\pi T)^2}
( \delta_{il}q_r+\delta_{ir}q_l+\delta_{lr}q_i  )\mbox{\boldmath$\sigma$}_{\alpha\beta}\psi^{\prime\prime}\left (\frac{1}{2}+\frac{1}{4\pi T\tau}   \right )
-\frac{\pi}{36}N_0v_F^2( \delta_{il}q_r+\delta_{ir}q_l)\mbox{\boldmath$\sigma$}_{\alpha\beta}T\sum_{n\ge0}\frac{1}{\tilde\varepsilon_n^3\varepsilon_n\tau},
\label {BB}
 \end{equation} 
 \end{widetext}
 where $\psi^{\prime\prime}(z)$ is the second derivative of the digamma function.

  The matrix of the fluctuation propagator  is given be Eq.(\ref{Fl}).
   The off-diagonal elements of this matrix can be omitted because they are proportional to higher order terms in components of vector ${\bf q}$:   $(\delta_{rs}+bq_rq_s)^{-1}= \delta_{rs} -bq_rq_s+\dots$. Performing integration over momenta  in Eq.(\ref{Fl}) we obtain at small $q$ and $\Omega$
    \begin{equation}
  L^{rs}({\bf q}, \Omega_k)=\frac{3}{N_0}\frac{\delta_{rs}}{\epsilon+a\Omega+\xi^2q^2}
  \label{L}
  \end{equation}
  Here, 
  \begin{equation}
  \epsilon=\ln\frac{T}{T_{c0}}+\psi\left (\frac{1}{2}+\frac{1}{4\pi T\tau}   \right )-\psi\left (\frac{1}{2}\right ).
 \end{equation}
 The critical temperature $T_c$ is 
  suppressed in respect to the temperature $T_{c0}$ of superfluid transition in pure helium and determined from the equation
  \begin{equation}
  \ln\frac {T_{c0}}{T_c}=\psi\left (\frac{1}{2}+\frac{1}{4\pi T_c\tau}   \right )-\psi\left (\frac{1}{2}\right ).
  \end{equation}
  The coefficient 
   \begin{eqnarray} 
a=\frac {\pi T}{(2\pi T)^2}\psi^{\prime}\left (\frac{1}{2}+\frac{1}{4\pi T\tau}   \right )~~~~~~~~\nonumber\\=\left\{
 \begin{array} {rc}&
 \frac{\pi}{8T}
  , ~~~~4\pi T\tau\gg 1,~~~~|\Omega|\ll4\pi T,\\
&\tau
,~~~~4\pi T\tau\ll 1,~~~~|\Omega|\tau\ll1.
 \end{array} \right.
  \end{eqnarray}
The first line here corresponds to the limit of weak scattering when the critical temperature is slightly suppressed by impurities
$(T_{c0}-T_c)/T_{c0}\approx(\pi/8T_{c0}\tau)\ll1$ and the typical frequencies of fluctuations $|\Omega|\approx (T-T_c)\ll T$. This is  quasi-static or classic fluctuation region. In respect to the second line one must remark that impurities completely  suppresses superfluidity at $\tau_c=\frac{\gamma}{\pi T_{c0}}$,
here $\gamma\approx 1.8$ is the Euler constant. Hence, $4\pi T\tau>\frac{4\gamma T}{T_{c0}}$ and for fulfillement of inequality
$4\pi T\tau\ll 1$ the temperature must be at least $1/4\gamma$ times lower than the critical temperature in pure helium. 
Still, at such low temperatures there are two different situations. First, this is again region of classic fluctuations $|\Omega|\tau\ll4\pi T\tau\ll 1$.
The second is the region of quantum fluctuations $4\pi T\tau\ll |\Omega|\tau\ll1$ when the frequencies of fluctuations exceed the temperature.

The coefficient 
 \begin{eqnarray}
 \xi^2=-\frac{1 }{40}\frac{v_F^2}{(2\pi T)^2}\psi^{\prime\prime}\left (\frac{1}{2}+\frac{1}{4\pi T\tau}   \right )\nonumber\\=\left\{
 \begin{array} {rc}
  \frac{7\zeta(3)}{20}\frac{v_F^2}{(2\pi T)^2}
  , ~~~~4\pi T\tau\gg 1,\\
 \frac{1}{10}v_F^2\tau^2
 ,~~~~4\pi T\tau\ll 1.
 \end{array} \right.
 \end{eqnarray}
 It is convenient to rewrite these expressions in terms of zero temperature coherence length $\xi_0=\frac{v_F}{2\pi T_c}$. Hence, at temperatures near $T_c$
  \begin{equation}
 \xi\simeq
 \left\{
 \begin{array} {rc}
 0.65\xi_0 
  , ~~~~4\pi T\tau\gg 1,\\
2T_c\tau\xi_0
 ,~~~~4\pi T\tau\ll 1.
 \end{array} \right.
 \end{equation}
Thus, unlike to the case of $s$-wave pairing  Ref.8, both in the clean case and in the dirty enough $T_c\tau\approx 1$ case    $\xi\approx\xi_0$.

Using this notation  the Eq. (\ref{BB}) acquires  the following form
\begin{eqnarray}
{\bf B}^{lr}_{i,\alpha\beta}=-\frac{2}{3}N_0\xi^2
( \delta_{il}q_r+\delta_{ir}q_l+\delta_{lr}q_i  )\mbox{\boldmath$\sigma$}_{\alpha\beta}\nonumber\\-\frac{2}{3}N_0\xi_1^2( \delta_{il}q_r+\delta_{ir}q_l)\mbox{\boldmath$\sigma$}_{\alpha\beta},
\label{BBBB}
\end{eqnarray}
where 
\begin{eqnarray}
 \xi^2_1=
 \left\{
 \begin{array} {rc}
  
  \frac{5\zeta(4)}{2^7}\frac{v_F^2}{(2\pi T)^24\pi T\tau}
  , ~~~~4\pi T\tau\gg 1,\\
\frac{1}{6}v_F^2\tau^2\ln\frac{e}{10\pi T\tau}
 ,~~~~4\pi T\tau\ll 1.
 \end{array} \right.\nonumber
 \end{eqnarray}

Let us first neglect the terms $\propto \xi_1^2$ that is true in low scattering limit $4\pi T\tau\gg1$.
Substituting Eqs.(\ref{L}) and (\ref{BBBB}) into Eq.(\ref{current}) and making use the analytical continuation from the discrete frequencies to the complex plane \cite{Aslamazov1968} we obtain linear in frequency term in the spin current 
\begin{widetext}
\begin{equation}
 {\bf j}_{i}(\omega)=-\frac{\omega}{4\pi i}\left(\frac{2}{3}N_0\xi^2\right)^2\int\frac{d^3q}{(2\pi)^3}\int d\Omega\frac{(L^R-L^A)^2}{2T\sinh^2\frac{\Omega}{2T}}(9q_iq_j+2q^2\delta_{ij})
  \mbox{\boldmath$\omega$}_j,
\end{equation}
\end{widetext}
where
\begin{equation}
 L^{R/A}({\bf q}, \Omega_k)=\frac{3}{N_0}\frac{1}{\epsilon\mp ia\Omega+\xi^2q^2}
\end{equation}
are  retarded and  advanced fluctuation propagators. Performing the integration in classic (static) $(T-T_c)\ll T$ limit  we have:
\begin{equation}
{\bf j}_{i}(\omega)=i\omega\frac{45}{16\xi\sqrt{\epsilon}}~
  \mbox{\boldmath$\omega$}_i.
  \label{cur}
\end{equation}

The quantum limit  ( in neglect of terms  $\propto \xi_1^2$) can be reached at low temperatures when  $ \epsilon\gg T\tau$ and the corresponding current expression is
\begin{equation}
{\bf j}_{i}(\omega)=i\omega\frac{5(\tau T)^2}{\xi\epsilon^{3/2}}~
  \mbox{\boldmath$\omega$}_i.
  \label{Q}
\end{equation}
Still, the temperature is limited from below by the critical temperature $T_c$ and the critical fluctuations in close vicinity of critical temperature are always   classical $ \epsilon\ll T\tau$ to displeasure of fans of quantum phase transitions.

Making use the Larmor theorem 
\begin{equation}
\gamma{\bf H}=\frac{\partial\mbox{\boldmath$\theta$}}{\partial t}=-i\omega\mbox{\boldmath$\theta$}
\end{equation}
where $\gamma=2\mu$ is the gyromagnetic ratio, $\mu$ is the magnetic moment of $^3$He atoms, one can rewrite the Eq. 
(\ref{cur}) for the fluctuation current as
\begin{equation}
{\bf j}_{i}^{fl}=-\frac{45}{8\xi\sqrt{\epsilon}}\mu\nabla_i{\bf H}.
\label{C}
\end{equation}

Thus, in low scattering limit $4\pi T\tau\gg1$, where  the negligence by  terms $\propto \xi_1^2$ is justified, Eqs.(22), (25) are correct.
Whereas
in  ultra low temperature region $4\gamma \frac{T}{T_c}<4\pi T\tau\ll 1$ we have $\xi_1^2=\frac{5}{3}\xi^2\ln\frac{e}{10\pi T\tau}$.
Hence,   in this temperature region in the classic (static) $(T-T_c)\ll T$  limit  the fluctuation current  given by Eqs.(22),(25)acquires factor 
$\propto\left (\ln\frac{e}{10\pi T\tau}\right)^2$. The same is true in respect of Eq. (23) in the quantum region $ \epsilon\gg T\tau$.

The sluggish logarithmic correction develops itself  only at very low temperatures, therefore it presents just academic interest.

\section{conclusion}

In conclusion it is reasonable to compare the spin current due to the Cooper pairs fluctuations with 
the diffusion current \cite{Mineev2018} determined by impurity scattering. The latter  in dimensional units is
\begin{equation}
{\bf j}_{i}^{dif}=-\hbar N_0D\mu\nabla_i{\bf H}.
\end{equation}
Here, $D=\frac{1}{3}\tau v_F^2$ is the spin diffusion coefficient. Thus, the ratio of two currents is
\begin{equation}
\frac{j^{fl}}{j^{dif}}=\frac{45}{8\hbar \xi N_0D}\frac{1}{\sqrt{\epsilon}}.
\end{equation}
In dense aerogel the diffusion coefficient can be small enough $D\approx10^{-3}~cm^2/sec$ \cite{Dmitriev2015}, the coherence length 
at ambient pressure \cite{Sauls2005} is $ \xi_0=
2\times 10^{-6}~ cm$, density of states  at ambient pressure \cite{Wheatley1975} is $N_0\approx 0.5(erg~ cm^3)^{-1}$.
Thus,
\begin{equation}
\frac{j^{fl}}{j^{dif}}\approx5\times 10^{-2}
\frac{1}{\sqrt{\epsilon}}.
\end{equation}
In respect of experimental detection of fluctuation spin current this result is not encouraging. 
But, it will be perhaps useful to exact  determination of temperature of transition at measurement of coefficient of diffusion near the critical temperature.

The temperature dependence of spin diffusion current in $^3$He  due to pair fluctuations (\ref{C}) is turned out  the same as the temperature dependence of paraconductivity of a normal 3D metal near transition to the s-wave superconducting state \cite{Aslamazov1968}. This not astonishing because the structure of the theory is similar despite some particular  features typical for $p$-wave pairing. The temperature dependence of the fluctuation spin diffusion current in quantum limit (\ref{Q})  also coincides with the temperature dependence of paraconductivity in $d$-wave superconductors in quantum limit found in Ref.11.

\begin{figure}[p]
\includegraphics
[height=.8\textheight]
{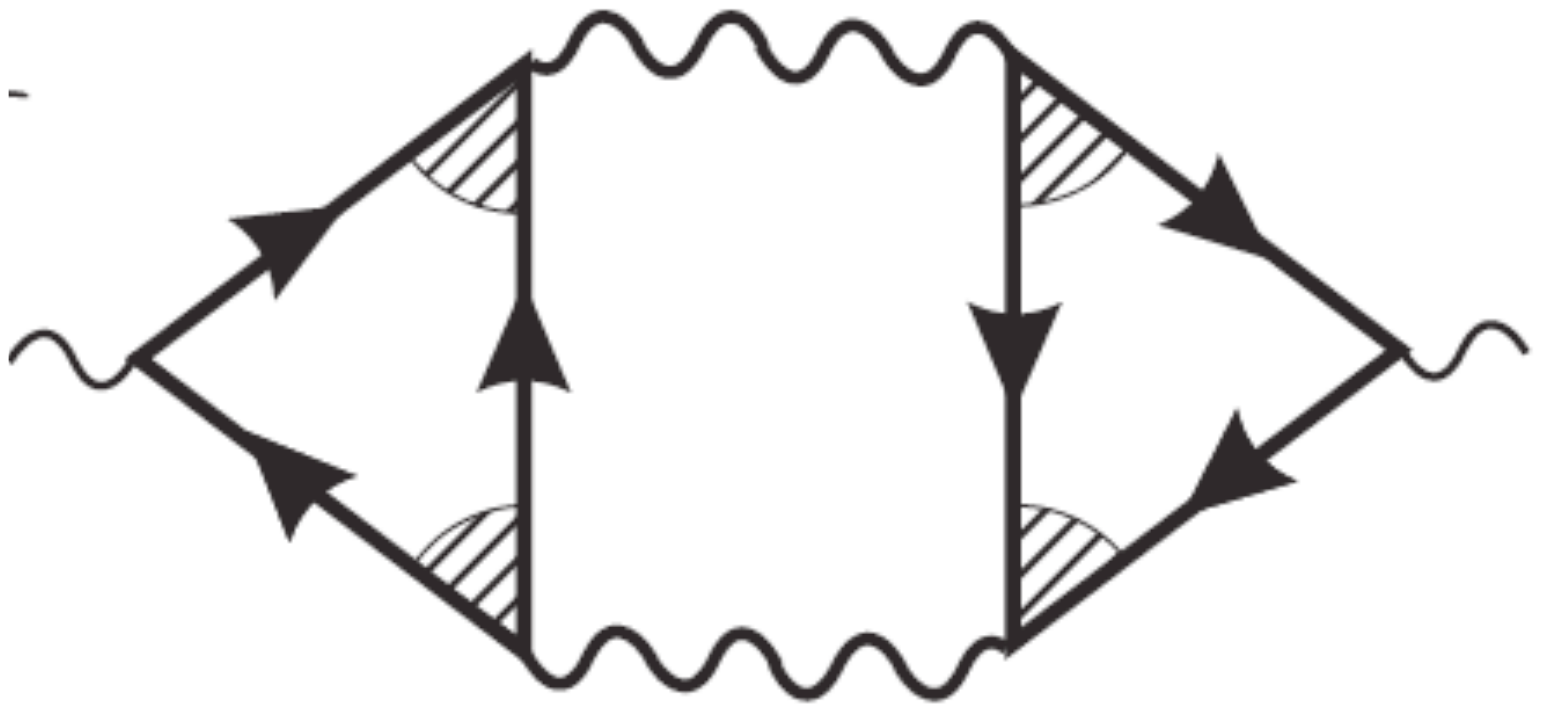}
 \caption{ 
 The Aslamazov-Larkin diagram.}
\end{figure}

\end{document}